\documentclass[useAMS,usenatbib]{mn2e}

\usepackage[active]{srcltx}
\usepackage{graphicx}
\usepackage{txfonts}
\usepackage{natbib}
\usepackage{multirow}
\usepackage{bigdelim}
\usepackage{longtable}
\usepackage[applemac]{inputenc}
\usepackage{rotating}
\usepackage{colours}

\newcommand{\xmm}{\textit{XMM-Newton}\ }

\newcommand{\kms}{km\, s$^{-1}\,$}

\def\gtrsim{\mathrel{\hbox{\rlap{\hbox{\lower4pt\hbox{$\sim$}}}\hbox{$>$}}}}
\def\ltsim{\mathrel{\hbox{\rlap{\hbox{\lower4pt\hbox{$\sim$}}}\hbox{$<$}}}}

\title[Plaskett's star]{Discovery of a magnetic field in the rapidly-rotating O-type secondary of the colliding-wind binary HD 47129 (Plaskett's star)\thanks{Based on spectropolarimetric observations obtained at the Canada-France-Hawaii Telescope (CFHT) which is operated by the National Research Council of Canada, the Institut National des Sciences de l'Univers (INSU) of the Centre National de la Recherche Scientifique of France, and the University of Hawaii, as well as on observations obtained using the Narval spectropolarimeter at the Observatoire du Pic du Midi (France), which is operated by the INSU. }}
\author[J.H. Grunhut et al.]{J.H. Grunhut\thanks{E-mail: jason.grunhut@gmail.com}$^{1,2}$, G.A. Wade$^1$, M. Leutenegger$^3$, V. Petit$^4$, G. Rauw$^5$, C. Neiner$^6$,  F. Martins$^7$,
\newauthor{D.H. Cohen$^8$, M. Gagn\'e$^4$, R. Ignace$^9$, S. Mathis$^{10}$, S.E. de Mink$^{11,12}$\thanks{Hubble Fellow.}, A.F.J. Moffat$^{13}$, }
\newauthor{S. Owocki$^{14}$, M. Shultz$^{1,2}$, J. Sundqvist$^{14}$, and the MiMeS Collaboration}\\
$^{1}$Department of Physics, Royal Military College of Canada, PO Box 17000, Stn Forces, Kingston, Ontario K7K 7B4, Canada \\
$^{2}$Department of Physics, Engineering Physics and Astronomy, Queen's University, Kingston, Ontario K7L 3N6, Canada \\
$^{3}$Department of Physics, University of Maryland, Baltimore County, 1000 Hilltop Circle, Baltimore, MD 21250, USA\\
$^{4}$Department of Geology \& Astronomy, West Chester University, West Chester, Pennsylvania, USA 19383\\
$^{5}$Institut d'Astrophysique et de G\'eophysique, Universit\'e de Li\`ege, Belgium\\
$^{6}$LESIA, UMR 8109 du CNRS, Observatoire de Paris, UPMC, Universit\'e Paris Diderot, 5 place Jules Janssen, 92195 Meudon Cedex, France\\
$^7$LUPM-UMR5299, CNRS \& Universit\'e Montpellier II, Place Eug\`ene Bataillon, F-34095, Montpellier, France\\
$^8$Department of Physics and Astronomy, Swarthmore College,  500 College Ave., Swarthmore, PA 19081, USA\\
$^{9}$Department of Physics and Astronomy, East Tennessee State University, Johnson City, TN 37663, USA\\
$^{10}$Laboratoire AIM Paris-Saclay, CEA/DSM-CNRS-Universit\'e Paris Diderot, IRFU/SAp Centre de Saclay, 91191, Gif-sur-Yvette, France\\
$^{11}$Space Telescope Science Institute, 3700 San Martin Drive, Baltimore, MD 21218, USA\\
$^{12}$Johns Hopkins University, 3400 North Charles St., Baltimore, MD 21218, USA\\
$^{13}$D\'epartement de physique, Universit\'e de Montréal C.P. 6128, Succursale Centre-Ville, Montr\'eal, QC H3C 3J7, Canada\\
$^{14}$Bartol Research Institute, University of Delaware, Newark, DE 19716, USA}

\begin{document}

\date{Accepted . Received , in original form }

\pagerange{\pageref{firstpage}--\pageref{lastpage}} \pubyear{2002}

\maketitle

\label{firstpage}

\begin{abstract}
We report the detection of a strong, organized magnetic field in the secondary component of the massive O8III/I+O7.5V/III double-lined spectroscopic binary system HD 47129 (Plaskett's star), in the context of the Magnetism in Massive Stars (MiMeS) survey. Eight independent Stokes $V$ observations were acquired using the ESPaDOnS spectropolarimeter at the Canada-France-Hawaii Telescope and the Narval spectropolarimeter at the T\'elescope Bernard Lyot. Using Least-Squares Deconvolution we obtain definite detections of signal in Stokes $V$ in 3 observations. No significant signal is detected in the diagnostic null ($N$) spectra. The Zeeman signatures are broad and track the radial velocity of the secondary component; we therefore conclude that the rapidly-rotating secondary component is the magnetized star. Correcting the polarized spectra for the line and continuum of the (sharp-lined) primary, we measured the longitudinal magnetic field from each observation. The longitudinal field of the secondary is variable and exhibits extreme values of $-810\pm 150$~G and $+680\pm 190$~G, implying a minimum surface dipole polar strength of $2850\pm 500$~G. In contrast, we derive an upper limit ($3\sigma$) to the primary's surface magnetic field of $230$~G. The combination of a strong magnetic field and rapid rotation leads us to conclude that the secondary hosts a centrifugal magnetosphere fed through a magnetically confined wind. We revisit the properties of the optical line profiles and X-ray emission - previously interpreted as a consequence of colliding stellar winds - in this context. We conclude that HD 47129 represents a heretofore unique stellar system - a close, massive binary with a rapidly rotating, magnetized component - that will be a rich target for further study.

\end{abstract}



\section{Introduction}

\begin{table*}
\caption{\label{log}Log of observations and magnetic diagnosis of Plaskett's star. Phases correspond to the orbital ephemeris of \citet{2008A&A...489..713L}, for which phase 0.0 denotes the time of primary conjunction. In the "Det?" column, ND=No Detection, DD=Definite Detection. Detection probabilities in Stokes $V$ ($P_{\rm V}$ and null $N$ $P_{\rm N}$) are given in percent. Longitudinal magnetic fields ($B_\ell$ for Stokes $V$, $N_\ell$ for the diagnostic null) correspond to the LSD profiles of the secondary component, using the mean Stokes $I$ profile of the secondary derived from the disentangled spectrum of \citet{2008A&A...489..713L} renormalised to the binary continuum. The longitudinal field detection significance ($z=B_\ell/\sigma$) is given for both Stokes $V$ ($z_{\rm B}$) and null $N$ ($z_{\rm N}$).}
\begin{center}
\begin{tabular}{p{1.1cm}cp{0.2cm}rrcrrrrrrrrrrr}\hline\hline
  \ \ \ \ HJD & Instrument    & &Phase & SNR &    Det?  & $P_{\rm V}$ & $P_{\rm N}$ & $B_\ell$ & $z_{\rm B}$ & $N_\ell$ & $z_{\rm N}$ \\
-2450000   &  &&   &&   &(\%)&(\%)&   (G) &  &(G)\\
                             \hline
5961.8434 &ESPaDOnS&&& &  &           & &               \\
5961.8731 &ESPaDOnS&&\multirow{-2}{*}{0.169}& \multirow{-2}{*}{2080}&  \multirow{-2}{*}{DD} & \multirow{-2}{*}{99.999}&\multirow{-2}{*}{91.693}&\multirow{-2}{*}{$530\pm 120$} &\multirow{-2}{*}{+4.6} &\multirow{-2}{*}{$-300\pm 110$} & \multirow{-2}{*}{-2.6}\\
5966.8313 &ESPaDOnS&&& &  &            &&            \\
5966.8610 &ESPaDOnS&&& &  &            &&            \\
5966.8911 &ESPaDOnS&&& &  &            &&            \\ 
5966.9229 &ESPaDOnS& \ldelim\} {-4}{0.1in} &\multirow{-4}{*}{0.518}&\multirow{-4}{*}{2428} & \multirow{-4}{*}{ND} &\multirow{-4}{*}{97.327}&\multirow{-4}{*}{17.014}& \multirow{-4}{*}{$260\pm 100$} &\multirow{-4}{*}{+2.6}& \multirow{-4}{*}{$80\pm 100$} &  \multirow{-4}{*}{+0.8} \\
5967.7528 &ESPaDOnS&&& &  &            &&            \\
5967.7825 &ESPaDOnS&&\multirow{-2}{*}{0.580}&\multirow{-2}{*}{1611} & \multirow{-2}{*}{DD} &\multirow{-2}{*}{100.00}&\multirow{-2}{*}{27.506}& \multirow{-2}{*}{$-810\pm 150$} &\multirow{-2}{*}{-5.5}& \multirow{-2}{*}{$50\pm 150$} &  \multirow{-2}{*}{+0.3} \\
5969.7301 &ESPaDOnS&&& &  &            &&            \\
5969.7598 &ESPaDOnS&&& &  &            &&            \\
5969.7900 &ESPaDOnS& \ldelim\} {-3}{0.1in}&\multirow{-3}{*}{0.719}&\multirow{-3}{*}{3000}& \multirow{-3}{*}{DD} &\multirow{-3}{*}{100.00}&\multirow{-3}{*}{57.525}& \multirow{-3}{*}{$510\pm 80$} &\multirow{-3}{*}{+6.4}& \multirow{-3}{*}{$10\pm 80$} & \multirow{-3}{*}{+0.1} \\
6000.4169 &Narval&&0.849&997 & ND &99.383&53.105& $-200\pm 250$ &-0.8& $400\pm 250$& +1.6\\ 
6001.3651 &Narval&&0.915&1072 & ND &86.539&47.268& $300\pm 230$ &+1.3& $110\pm 240$ & +0.5 \\
6010.3586 &Narval&&0.540&601 & ND &32.638&89.268& $250\pm 410$ &+0.6& $290\pm 410$ &  +0.7\\
6012.3439 &Narval&&0.677&1217 & ND &96.365&86.136& $680\pm 190$ &+3.6& $-50\pm 190$ & -0.3\\

\hline\end{tabular}
\end{center}
\end{table*}

%


HD 47129 (Plaskett's star) is an optically bright ($V=6.06$), high-mass \citep[$M_{\rm tot}\sin^3 i=92.7\pm 2.7~M_{\odot}$;][]{2008A&A...489..713L}, short-period \citep[$P_{\rm orb}=14.396257\pm 0.00095$~d;][]{2008A&A...489..713L}, approximately circular \citep{1987Obs...107...68S} non-eclipsing O+O SB2 system. It is considered to be a probable member of Mon OB2, a large association located at a heliocentric distance of $1.4-1.7$~kpc \citep{2007AJ....134.1368C} containing two or three subgroups of OB stars and stellar aggregates \citep{1979Ap&SS..66..191S,2005AJ....130..721L}. These components are characterized by a relatively large range of characteristic ages, from 0.2–0.6 Myr \citep[the dynamical age of NGC 2244, the youngest component of the association;][]{1967ApJ...147..965M} to 2 Myr \citep[the main sequence turnoff age of NGC 2244;][]{2002AJ....123..892P} to 4 Myr \citep[the dynamical age of the H~{\sc i} shell of the Rosette;][]{2007AJ....134.1368C} to 20-25 Myr \citep{1976ApJ...210...65T} for the most evolved stars in the association.  

According to the analysis of \citet{2008A&A...489..713L}, the system is composed of an O8III/I primary (HD 47129A1) of mass $(45.4\pm 2.4)\sin^3 i~M_\odot$ and $v\sin i=60-75$~\kms, and an O7.5V/III secondary (HD 47129A2) of mass $(47.3\pm 0.3)\sin^3 i~M_\odot$ and $v\sin i=230-310$~\kms. Despite their very similar masses, the primary appears to be $\sim 2\times$ brighter than the secondary in the optical \citep{2008A&A...489..713L}. The system inclination $i=71\pm 9\degr$ was estimated by \citet{1978PASP...90..163R} using linear polarimetry. \citet{1978PASP...90..163R} and \citet{1992ApJ...385..708B} noted that an inclination significantly larger than $\sim 70\degr$ would result in eclipses (which are not observed), while an inclination significantly smaller would imply improbably large masses of the components. Therefore, with a total mass of approximately $112~M_\odot$, HD 47129 is one of the most massive known O-type binaries. 

The system is clearly chemically peculiar: according to \citet{2008A&A...489..713L}, the primary is strongly N enhanced (16 times solar) and C depleted (3\% of solar), while the secondary is N depleted (20\% of solar) and He overabundant (1.5 times solar). The peculiar chemistry, in combination with the mass/luminosity mismatch and rapid rotation of the secondary, has led investigators \citep{1992ApJ...385..708B, 2008A&A...489..713L} to speculate that Plaskett's star is a post Roche lobe overflow (RLOF) system.

The system was first identified as an X-ray emitter by the ROSAT All-Sky Survey \citep{1996A&AS..118..481B}. \citet{2006MNRAS.370.1623L} reported analysis of \xmm observations of Plaskett’s star, revealing the system to be a hard, luminous and variable X-ray emitter ($kT_{\rm max} \simeq 1.4$~keV, $\log L_{\rm x}/L_{\rm bol} = -6.0$). These observations support the historical interpretation \citep[e.g.][]{1992ApJ...396..238W} of the system as a colliding-wind binary (CWB).


HD 47129 was observed in the context of the Magnetism in Massive Stars (MiMeS) survey of magnetic fields in Galactic OB stars. This target was flagged as a probable Zeeman detection following preliminary analysis via the MiMeS pipeline. The primary goal of this paper is to report the firm detection of a strong magnetic field in the secondary component of the system. This discovery leads directly to a re-examination of the optical and X-ray properties of Plaskett's star as a Magnetically-Confined Wind Shock (MCWS) system or CWB/MCWS hybrid. The latter classification would make HD 47129 unique among the known massive magnetic stars.


\section{Observations}

\begin{figure*}
\centering
\includegraphics[width=15cm]{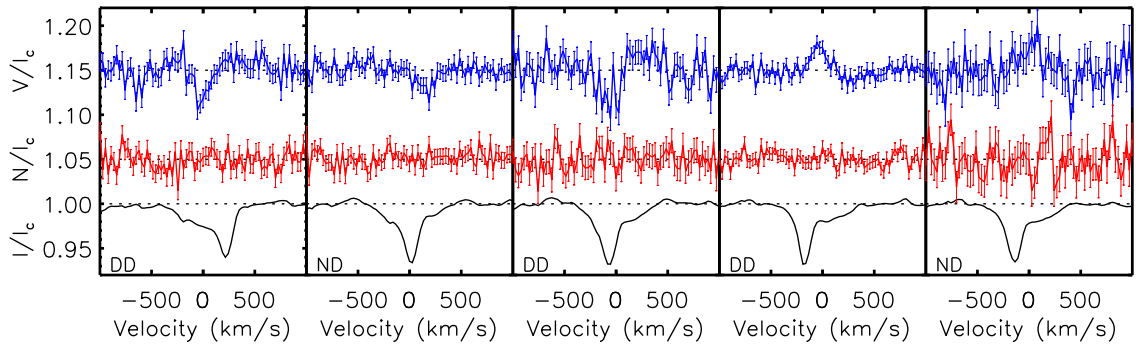}
\caption{\label{LSD1}Stokes $I$ (bottom), $V$ (top) and diagnostic null $N$ (middle) LSD profiles corresponding to the highest SNR polarized spectra (From left: JDs 2455961 (DD), 5966 (ND), 5967 (DD), 5969 (DD) and 6012 (ND)). No significant signal is detected in $N$. $V$ and $N$ have been scaled (by $300\times$) and shifted for display purposes. Note the complexity and variability of Stokes $I$.}
\end{figure*}

Fifteen high resolution ($R\simeq 65\,000$) spectropolarimetric (Stokes $I$ and $V$) observations of HD 47129 were collected with ESPaDOnS at the Canada-France-Hawaii Telescope (CFHT) and Narval at the T\'elescope Bernard Lyot (TBL) between 3 Feb and 25 Mar 2012. Each spectropolarimetric sequence consisted of four individual subexposures, each of 600 s duration (for ESPaDOnS) or 1\,200 s duration (for Narval), taken in different configurations of the polarimeter retarders. From each set of four subexposures we derived Stokes $I$ and Stokes $V$ spectra in the wavelength range $3\,670-10\,000$~\AA\ following the double-ratio procedure described by \citet{1997MNRAS.291..658D}, ensuring in particular that all spurious signatures are removed to first order. Diagnostic null polarization spectra (labeled $N$) were calculated by combining the four subexposures in such a way that polarization cancels out, allowing us to verify that no spurious signals are present in the data \citep[for more details on the definition of $N$, see][]{1997MNRAS.291..658D}. All frames were processed using the automated reduction package Libre ESpRIT \citep{1997MNRAS.291..658D}. 

All spectra acquired during a single night of observation were co-added following reduction to increase the signal-to-noise ratio (SNR). This ultimately resulted in 8 independent spectropolarimetric observations (i.e. 4 co-added ESPaDOnS spectra and 4 individual Narval spectra). The peak SNRs per 1.8 \kms\  pixel in these spectra range from 600-3000. The log of observations is reported in Table~\ref{log}.

\section{Detection of a magnetic field in Plaskett's star}
\label{stokesi}

As described by e.g. \citet{2008A&A...489..713L}, the optical spectrum of HD 47129 is extremely complex. Due to the broad lines of the secondary, spectral lines of the two components are never well-separated in velocity during the $\pm 200$~\kms amplitude radial velocity (RV) orbits of the two stars. Many lines exhibit multiple, variable (in position and profile shape) emission and absorption components. These spectral characteristics make a detailed diagnosis and interpretation of the magnetic properties of the system very challenging. In this first analysis, we employ Least-Squares Deconvolution \citep[LSD;][]{1997MNRAS.291..658D} and take advantage of the disentangled spectra of the components computed by \citet{2008A&A...489..713L} to aid in the interpretation of our results.

Examination of the best-quality co-added Stokes $V$ spectra reveals no strong evidence of Zeeman signatures in individual spectral lines of the primary or secondary (although, as described above, the complexity of the spectrum impedes such an identification). To improve our detection capabilities, LSD was applied to the reduced spectra. To begin, we adopted a line mask based on the LTE synthetic spectrum of an O8 giant. As in the case of other magnetic O-type stars with complex spectra \citep[e.g.][]{2006MNRAS.365L...6D, 2010MNRAS.407.1423M}, we found that a relatively small subset of lines {(12)} contribute usefully to the detection of a Zeeman signature. We ultimately arrived at an optimal line mask, yielding the most significant Zeeman detections, with content similar to the mask employed by \citet{2006MNRAS.365L...6D} for HD 191612. For more detail regarding line mask optimization for O-type stars with complex spectra, the reader is invited to consult, e.g. \citet{2011MNRAS.416.3160W,2012arXiv1206.2834W}.

Extraction of LSD profiles from the 8 spectropolarimetric observations using the adopted mask yields 3 definite detections and 5 non-detections of significant signal in the Stokes $V$ LSD profiles \citep[according to the detection criteria described by][]{1997MNRAS.291..658D}. In no case is any significant signal detected in the diagnostic null. The LSD profiles corresponding to the five highest-SNR spectra - i.e. 3 definite detections as well as two higher-SNR non-detections - extracted on a velocity grid spanning 2000~\kms sampled at 27~\kms, are illustrated in Fig.~\ref{LSD1}. Note that while a majority of our observations correspond formally to non-detections, the Stokes $V$ profiles of most observations are visibly more structured (and exhibit larger detection probabilities) than the corresponding $N$ profiles (see, for example, the second and fifth panels of Fig.~\ref{LSD1}, as well as Table 1).

A first examination of the LSD profiles in Fig.~\ref{LSD1} reveals that the Stokes $V$ Zeeman signatures are generally much broader than the sharp, easily identified Stokes $I$ profile of the primary star. This strongly suggests that the Stokes $V$ profile has its origin in the spectrum of the (broad-lined) secondary star. Unfortunately, while the primary's contribution to the LSD Stokes $I$ profiles is distinctive, the position and range of the broad secondary profile is more difficult to determine. To aid us in this task, we used the disentangled spectra of the primary and secondary computed by \citet{2008A&A...489..713L}. For each of the 8 observed orbital phases, we applied appropriate RV shifts to the disentangled spectra corresponding to the orbital solution of \citet{2008A&A...489..713L}. We then extracted the Stokes $I$ LSD profiles from these shifted spectra in the same manner as was applied to the ESPaDOnS and Narval Stokes $I$ spectra. 

\begin{figure*}
\centering
\includegraphics[width=15cm]{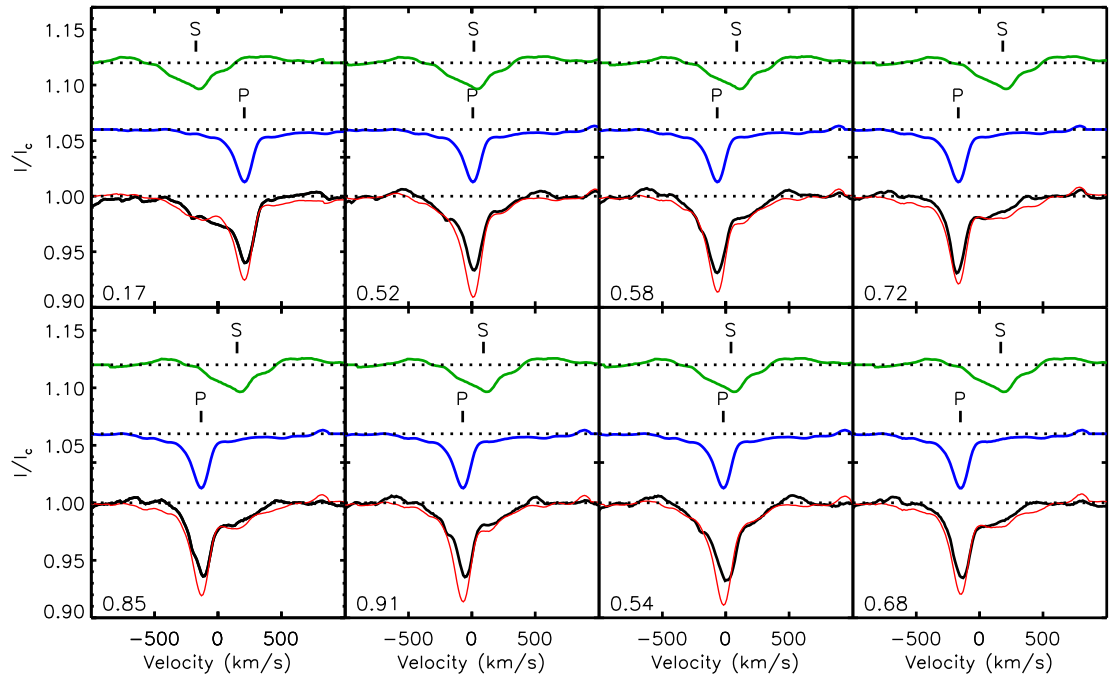}
\caption{\label{LSD2}Observed and modelled Stokes $I$ LSD profiles of HD 47129 (in chronological order, left to right and top to bottom). {\em Top row in each frame:}\ LSD profile of the secondary shifted to the predicted secondary RV at the time of observation. {\em Middle row in each frame:}\ LSD profile of the primary shifted to the predicted primary RV at the time of observation. {\em Bottom row in each frame, thick (black) line:} Observed LSD profile of HD 47129. {\em Bottom row, thin (red) line:}\ model LSD profile obtained from weighted and renormalised addition of the LSD profiles of the individual components extracted from the disentangled spectra. Orbital phase is indicated in the lower left corner of each frame.}
\end{figure*}

In principle, these LSD profiles should be representative of the individual components' LSD profiles at the observed orbital phases, assuming minimal intrinsic variability of the stars. To check the limitations of this assumption, we computed the binary LSD profiles expected at the observed phases by weighting the individual profiles according to the optical brightness ratio \citep[$L_{\rm pri}/L_{\rm sec}=1.9$;][]{2008A&A...489..713L}, summing them and renormalising them to the combined binary continuum. Comparing these to the observed binary LSD profiles (from the ESPaDOnS/Narval spectra; Fig.~\ref{LSD2}) we find an acceptable agreement. We therefore conclude that while the primary and secondary are variable, the magnitude of this variability is not sufficient to introduce fundamental uncertainties into the following analysis. This conclusion is consistent with that of \citet{1992ApJ...385..708B}, who report that "the secondary contribution to the cross-correlation profiles is relatively constant with orbital phase."

Comparing the observed Stokes $V$ profiles with the position and width of the primary's mean Stokes $I$ profile confirms our initial assessment: the primary spectrum is not the origin of the detected Zeeman signatures. On the other hand, the position and breadth of the secondary's mean Stokes $I$ profile agree well with the Stokes $V$ profiles at all phases. The data therefore support the hypothesis that the secondary component of HD 47129 is a rapidly rotating, magnetized O-type star.

\section{Magnetic field characteristics}

\subsection{Measurement of the longitudinal field}
\label{magnetic}

Having concluded that the secondary component is responsible for the observed Stokes $V$ Zeeman signature, we wish to constrain the magnetic strength and geometry of the star at the level of the photosphere (where the lines employed in the LSD mask are principally formed). In the following we assume that the primary star contributes negligibly to the observed circular polarization. While this assumption is consistent with the observations, we cannot strongly test it given that the primary's profile is always blended with that of the secondary. On the other hand, it does not appear that the primary contributes to Stokes $V$ at a level comparable to the secondary.

The simplest quantitative measure of the stellar magnetic field from circular polarization measurements is the mean longitudinal magnetic field (or simply longitudinal field), computed from the first-order moment of the Stokes $V$ profile about its velocity centre-of-gravity $v_{\rm 0}$ \citep{1997MNRAS.291..658D, 2000MNRAS.313..851W} according to the equation:

\begin{equation}
B_\ell=-\frac{2.14\times 10^{11}}{\lambda\, z\, c}\ \frac{{\displaystyle \int (v-v_{\rm 0}) V(v)\ dv}}{\displaystyle {\int [1-I(v)]\ dv}}.
\end{equation}

In Eq. (1) $V(v)$ and $I(v)$ are respectively the continuum-normalised Stokes $I$ and $V$ profiles as a function of velocity, $v$. The wavelength $\lambda$ is expressed in nm and the longitudinal field $B_\ell$ is in gauss. The wavelength and Land\'e factor $z$ correspond to the weighting factors used in calculation of the LSD profiles {(500 nm and 1.2, respectively)}. In our mask, the mean SNR-weighted wavelength and Land\'e factor were 511 nm and 1.11, respectively.

Note that the integral in the denominator of Eq. (1) represents the equivalent width of the Stokes $I$ profile (in velocity units). To infer the longitudinal magnetic field of the secondary as a single star, we therefore require the secondary's line profile independent of the primary's. We attempted to subtract the primary profile from the combined profile at each phase using the primary's LSD profile from the disentangled spectrum. However, the primary's line is sufficiently sharp and deep that relatively small variability or errors in RV seriously distort the resultant secondary profile. Ultimately, we decided to replace the binary Stokes $I$ profile with the secondary's LSD profile from the disentangled spectrum. The accuracy of this choice is limited in two ways. First, the secondary profile is variable \citep{2008A&A...489..713L}; it is therefore likely that its equivalent width also varies. Secondly, the profile shows important emission in its wings. In all likelihood this emission is formed in the (probably magnetically-confined) wind, above the photosphere of the star, where the magnetic field is much weaker. Consequently, we assume that the emission does not contribute significantly to the Stokes $V$ profile (this assumption is revisited in Sect. 7). On the other hand, it reduces the equivalent width of Stokes $I$ if included in the range of velocity used to evaluate Eq. (1). 

\begin{figure*}
\centering
\includegraphics[width=15cm]{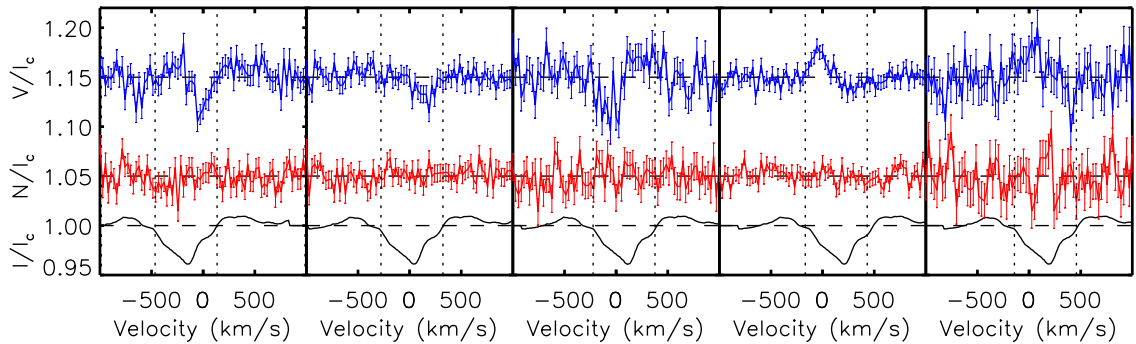}
\caption{\label{LSD3}Stokes $I$ (bottom), $V$ (top) and diagnostic null $N$ (middle) LSD profiles corresponding to selected observations (as in Fig.~\ref{LSD1}) used for the calculation of the longitudinal field of HD 47129. $V$ and $N$ have been scaled and shifted for display purposes. The observed binary Stokes $I$ profile has been replaced by a shifted and renormalised LSD profile of the secondary star, computed using the disentangled spectrum of \citet{2008A&A...489..713L}. The integration bounds used to evaluate Eq. (1) are illustrated with vertical dotted lines.}
\end{figure*}

Therefore to evaluate Eq. (1) we used the LSD profile from the disentangled secondary spectrum, shifted to the RV appropriate to each observation, and renormalised to the continuum of the real system \citep[using the optical brightness ratio of][]{2008A&A...489..713L} (Alternatively, we could have renormalised the Stokes $V$ profile to the continuum of the secondary. It can be shown that for the calculation of $B_\ell$ these operations are equivalent.). Eq. (1) was evaluated in the velocity range $\pm 300$~\kms symmetric about $v_{\rm 0}$. The velocity range was selected based on the reported $v\sin i$ of the secondary, as well as visual examination of the Stokes $I$ and $V$ profiles. We find that this range corresponds well to the central absorption region of the line profile. We recognize that emission may partially infill the absorption line, but we have no reasonable way to estimate the importance of such infilling. However, if our hypothesis that the emission results from a rigidly-rotating centrifugal magnetosphere is correct, then emission will be confined to velocities larger than $v=\pm v\sin i$, and will therefore not significantly infill the line profile. A selection of the LSD profile sets used for determination of the longitudinal field are illustrated in Fig.~\ref{LSD3}. In this figure we show the Stokes $V$ and null $N$ profiles (identical to those in Fig.~\ref{LSD1}), and the shifted and renormalised Stokes $I$ LSD profile of the secondary derived from the disentangled spectrum. The integration bounds used to evaluate Eq. (1) are illustrated with vertical dotted lines.

\begin{figure}
\centering
\includegraphics[angle=-90,width=8cm]{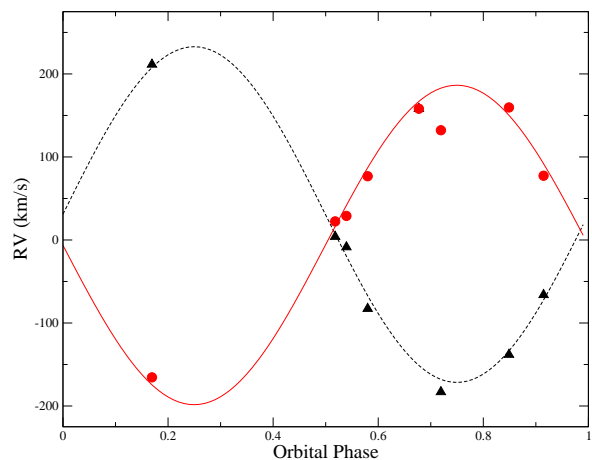}
\caption{\label{RVs}{\em Triangles:}\ Radial velocities of the primary component measured from ESPaDOnS and Narval Stokes $I$ LSD profiles. {\em Circles:}\ Centres-of-gravity of the Stokes $V$ LSD profiles. The dashed and solid lines represent respectively the theoretical RV variations of the primary and secondary components according to the orbital model of \citet{2008A&A...489..713L}. Clearly the Stokes $V$ profiles follow the secondary's RV variation, further establishing the association of the magnetic field with the secondary component.}
\end{figure}

A useful product of the calculation of the longitudinal magnetic field is the determination of the centre-of-gravity $v_{\rm 0}$ of Stokes $V$ at each phase. These measurements provide a quantitative verification of our tentative conclusion that the magnetic field is associated with the secondary component. As illustrated in Fig.~\ref{RVs}, the $v_{\rm 0}$ measurements track the RV of the secondary faithfully. In fact, these measurements provide an independent verification of the RV variation of the secondary that is independent of the disentangling procedures employed in past studies. Clearly they do not track the primary's RV curve well at all. 

It is therefore clear that the dominant contribution to Stokes $V$ comes from the secondary. Nevertheless, a weak contribution to Stokes $V$ - even above the level of the noise - could possibly come from the primary, and remain hidden in the stronger Stokes $V$ profile of the secondary.  Due to the primary's sharper spectral lines, a magnetic field is intrinsically more easily detectable in this star. Based on the ratio of projected rotational velocities of the two stars, we would roughly expect the primary's Stokes $V$ signature to have an amplitude equal to that of the secondary if its longitudinal magnetic field were one-fourth as large as the secondary's (i.e. $(v\sin i)_{\rm 2}/(v\sin i)_{\rm 1}=70/305\simeq 0.25$). Based on our analysis, we consider such a signal to be ruled out at 3$\sigma$ confidence. Taking into account the optical brightness ratio of the two stars, this yields a $3\sigma$ upper limit on the longitudinal field of the primary on JD 2455969 (i.e. when the secondary's field was detected with the greatest significance) of about 70~G.


\subsection{Variability of the longitudinal field}

The longitudinal field measurements of the secondary are reported in Table 1. They are observed to vary from an extreme positive value of $+680\pm 190$~G to an extreme negative value of $-810\pm 150$~G. The variation is significant at about 6.2$\sigma$, and the change of the sign is reflected in the changing shape of the Stokes $V$ profile. The longitudinal field measured on consecutive or near-consecutive nights - in particular on 9/10/12 Feb - is observed to be significantly different, indicating that the variability timescale is rather short, likely of order 1~d.

We have examined the period content of the longitudinal field measurements using the FFT and clean procedures as implemented in {\sc clean-NG} \citep{2009A&A...506..133G}. The only significant period present in the 8 longitudinal field measurements corresponds to $3.6\pm 0.1$~d, and yields an approximately sinusoidal variation of $B_\ell$, with a reduced $\chi^2$ of 1.37. (In contrast, folding the data with the orbital period yields a substantially worse fit, with reduced $\chi^2=7.46$.) However, given the relatively low SNR of the variation and the small number of data points, the reliability of this period is unclear. However, it is still instructive to compare it to the expected rotational periods of the stars.

Rotational periods of the components have been computed using radii inferred from the temperatures and luminosities reported by \citet{2008A&A...489..713L}. The uncertainties on the luminosities derived by \citet{2008A&A...489..713L} do not account for the uncertain distance to Mon OB2. To account for the distance modulus uncertainty, we adopt a luminosity uncertainty 0.1 dex. We note that this calculation does not take into account any rotational effects (e.g. oblateness or gravity darkening), nor does it include any mutual gravitational deformation of the stars. The projected rotational velocities derived by \citet{2008A&A...489..713L} using the Fourier technique on He~{\sc i} and He~{\sc ii} lines range from $60-75$~\kms for the primary and $230-310$~\kms for the secondary. For the purposes of our calculations, for the primary we adopt $v\sin i=70\pm 10$~\kms based on the results of \citet{2008A&A...489..713L}, but weighted to slightly higher velocity taking into account the result of \citet{1992ApJ...385..708B}. For the secondary, we adopt $305\pm 15$~\kms. Here we have taken into account only the results for He~{\sc ii} lines by \citet{2008A&A...489..713L}, since the He~{\sc i} lines are visibly affected by emission and result in substantially lower $v\sin i$ values. The adopted $v\sin i$ of the secondary is consistent with the result of \citet{1992ApJ...385..708B}. 

For the convenience of the reader, the relevant data are summarized in Table 2. We derive the radii of the components: $14.1\pm 3.3~R_\odot$ for the primary, $10.7\pm 2.6~R_\odot$ for the secondary. We obtain rotational periods $P_{\rm rot, 1}/\sin i=10.2\pm 4.0$~d and  $P_{\rm rot, 2}/\sin i=1.8\pm 0.5$~d. If we assume that the rotational and orbital angular momenta are aligned\footnote{While this appears to be a reasonable assumption for HD 47129 (see Sect. 7), other close, high-mass binaries exist for which this may not be the case \citep{2005ApJ...623.1092V,2006ApJ...640..995V}.} and adopt an inclination $i=71\degr$ \citep{1978PASP...90..163R} , then $P_{\rm rot, 1}=9.6\pm 3.7$~d and  $P_{\rm rot, 2}=1.7\pm 0.5$~d. The longest possible secondary rotational period consistent with the derived radii and published $v\sin i$ (i.e. for $i=90\degr$) is $1.8+0.5=2.3$~d. Therefore the 3.6~d period inferred from the FFT is not straightforwardly compatible with rotational modulation by the secondary.  


\citet{2011A&A...525A.101M} analyzed 34 days of continuous CoRoT photometry of Plaskett's star. The two most significant periods detected correspond to the orbital period of $14.4$~d, and 1.22~d. They report no immediate explanation for the 1.22~d period (although we do note that it is approximately one-third of 3.6~d), but conclude that this is not likely to be related to the rotational period of either star. This could only be the base if the stellar rotation axis were inclined to our line-of-sight by $43\degr$, in which case this value would be consistent with the inferred radius and $v\sin i$ of the secondary. On the other hand, as discussed in Sect. 1, if the rotation and orbital axes are aligned, such an inclination would imply masses of the components significantly (2.6 times) larger than those adopted by \citet{2008A&A...489..713L}.

Two additional periods, 2.51~d and 2.71~d, are reported by \citet{2011A&A...525A.101M}. These are similar to the period of variation of the H$\alpha$ wings of $P=2.86$~d reported by \citet{1992ApJ...396..238W}, and somewhat longer than the longest admissible secondary rotation period. Otherwise we are not yet able to offer any useful interpretation of them.

Given the relatively large amplitude of the longitudinal magnetic field of HD 47129A2, we are confident that further spectropolarimetric observations will yield a direct determination of the secondary's rotation period.






\subsection{Magnetic strength and geometry}

Although the available magnetic data are insufficient to develop a unique model of the magnetic geometry of HD 47129A2, it is reasonable to assume (similar to other magnetic O stars) that HD 47129A2 is an oblique rotator, and (given the large value of the longitudinal field and its strong variability) that the magnetic field has a relatively simple structure - i.e. it is dominated by the dipole component. In this case we can use Eq. (1) of \citet{1967ApJ...150..547P} to estimate the minimum possible polar strength of the surface dipole. Using a limb-darkening coefficient of 0.3 and the peak longitudinal field $|B_\ell|_{\rm max}=810\pm 150$~G, we compute a minimum polar field strength of the dipole $B_{\rm d}=2850\pm 500$~G.  

While we are not able to unambiguously determine the period of variation of the longitudinal field, it is clearly detected to vary and to change sign. This indicates that both magnetic hemispheres are visible during the star's rotation. This implies that at least one of the two angles specifying the magnetic field geometry (the inclination $i$ of the rotation axis to our line-of-sight, and the obliquity $\beta$ of the magnetic axis relative to the rotation axis) is relatively large, and that $i+\beta>90\degr$. 

Applying similar reasoning to our upper limit on the longitudinal field of the primary, we obtained a $3\sigma$ upper limit on the primary's surface dipole component of approximately $250$~G, i.e. more than $10\times$ smaller than that detected in the secondary.


\section{The wind-magnetic field and wind/wind interactions}
\label{magnetosphere}

\begin{table}
\centering
\caption{\label{param_summary}Summary of physical properties of HD 47129 used to compute component rotational periods. $T_{\rm eff}$, $\log L_*$ and $v\sin i$ are values reported by \citet{2008A&A...489..713L} (except that the luminosity uncertainties have been increased from $0.04$ to $0.1$ dex, as described in the text). $R_*$ and $P_{\rm rot}$ are calculated from those values.}
\begin{tabular}{lllll}
\hline
 & Primary & Secondary\\
 \\
$T_{\rm eff}$ (K) & 33\,500 $\pm$ 2000 & 33 000 $\pm$ 2000 \\
$\log (L_*/L_\odot)$ & $5.35\pm 0.1$ & $5.09\pm 0.1$   \\
$R_{*}$ ($R_\odot$) & $14.1\pm 3.3$ & $10.7\pm 2.6$ \\
$v\sin i$ (km\,s$^{-1}$) & $70\pm 10$ & $305\pm 15$  \\
$P_{\rm rot}/\sin i$ (d) & $10.2\pm 4.0$ & $1.8\pm 0.5$ \\
\hline\hline
\end{tabular}
\end{table}

As discussed by \citet{2002ApJ...576..413U}, the global competition between a magnetic field and stellar wind can be characterized by the so-called wind magnetic confinement parameter $\eta_* \equiv B^2_\mathrm{eq}R^2_* / \dot{M}v_\infty$, which depends on the star's equatorial surface field strength ($B_\mathrm{eq}=B_{\rm d}/2$), radius ($R_*$), and wind terminal momentum ($\dot{M}v_\infty$). For a dipolar field, one can identify an Alfv\'en radius $R_{\rm Alf}\simeq\eta_*^{1/4}~R_*$, representing the radial extent of strong magnetic confinement of the wind. 
In the equatorial plane, above $R_{\rm Alf}$, the wind dominates and stretches open all field lines. But below $R_{\rm Alf}$, the wind material is trapped by closed field line loops. In the case of a rapidly rotating star like HD 47129A2, centrifugal support results in a long residence time of confined plasma located above the Kepler co-rotation radius $R_{\rm Kep}\equiv(GM/\omega^2)^{1/3}$, where $\omega$ is the rotational frequency of the star (i.e. $\omega=v_{\rm rot}/R_*$). 

Although the primary's mass loss rate and wind terminal speed were determined observationally by \citet{2008A&A...489..713L}, the secondary's were not. In fact, \citet{2008A&A...489..713L} remark on the poor match of their CMFGEN spectra to optical and UV spectra, and the puzzling absence of the classical wind lines in the secondary's UV spectrum. 

To estimate the wind momentum of the secondary, we used the adopted parameters given in Table~\ref{param_summary}, and masses of $54~M_\odot$ for the primary and $56~M_\odot$ for the secondary (i.e. the projected masses cited in Sect. 1 evaluated for an orbital inclination of $71\degr$). While outside the scope of this paper, we note that the spectroscopic masses determined using the inferred temperature and luminosity are significantly (a factor of two for the primary, and 4 for the secondary) smaller than those determined dynamically. This is a serious discrepancy that may be tied both the the peculiar spectra and to the exotic history of the system, discussed further in Sect. 7. Whatever its origin, this discrepancy demands urgent attention.

We determined the theoretical wind terminal velocity and the mass-loss rate according to \citet{2000A&A...362..295V,2001A&A...369..574V}. We obtained $v_{\infty,{\rm 2}}=3500$\,km\,s$^{-1}$ and $\dot{M}_{\rm 2}=7.1\times 10^{-8}$\,M$_\odot$\,yr$^{-1}$. Although the calculated terminal speed is higher than that measured for the primary \citep[$v_{\infty, {\rm 1}}=2450$~\kms;][]{2008A&A...489..713L}, the mass loss rate is lower \citep[$\dot M_{\rm 1}=1.7\times 10^{-7}$\,M$_\odot$\,yr$^{-1}$][]{2008A&A...489..713L}. (For comparison, similar calculations for the primary yield $v_{\infty, {\rm 1}}=2980$~\kms and $\dot M_{\rm 1}=2.9\times 10^{-7}$\,M$_\odot$\,yr$^{-1}$, which are in reasonable agreement with the measured values.) From a physical perspective, $\dot M$ of the secondary is expected to be lower because the star is less luminous than the primary, but $v_\infty$ is larger because the escape velocity (determined by the surface gravity, $GM_*/R_*^2$) is larger.

These wind characteristics combined with the estimate of the surface magnetic field strength imply $\eta_*=680$: the magnetic field strongly confines the wind, dominating it dynamically to $R_{\rm Alf}=5.4~R_*$ in the magnetic equatorial plane. If we underestimated the wind momentum of the secondary by a factor of e.g. two, this would result in an overestimation of  $R_{\rm Alf}$ by $1.3~R_*$. We conclude that the uncertainty in $R_{\rm Alf}$ is dominated by the systematics in the wind momentum calculation, and not the formal errors of the stellar parameters. In addition, we recall that $B_{\rm d}$ represents a lower limit to the magnetic field strength, and is influenced by any systematic uncertainty in the optical brightness ratio. 


HD 47129A2 is the most rapidly-rotating known magnetic O-type star. We therefore expect rotation to have an important role in determining the structure of its magnetosphere. Using the values in Table~\ref{param_summary} and adopting $\sin i=1$ (i.e. assuming the rotation is the slowest permitted by $v\sin i$) we compute the Kepler co-rotation radius of the secondary, $R_{\rm Kep}\ltsim 2.2~R_*$. That is already below our best estimate of the lower limit on $R_{\rm Alf}$. This implies that HD 47129A2 hosts a region, between $R\simeq 2.2~R_*$ and $R\simeq 5.4~R_*$ (i.e. from 1.2-4.4~$R_*$ above the stellar surface) in which the wind is both magnetically confined and centrifugally supported. Thus we conclude that the secondary hosts a centrifugal magnetosphere - {\em the first such configuration amongst the known magnetic O-type stars.}

\begin{figure}
\centering
\includegraphics[angle=-90,width=9cm]{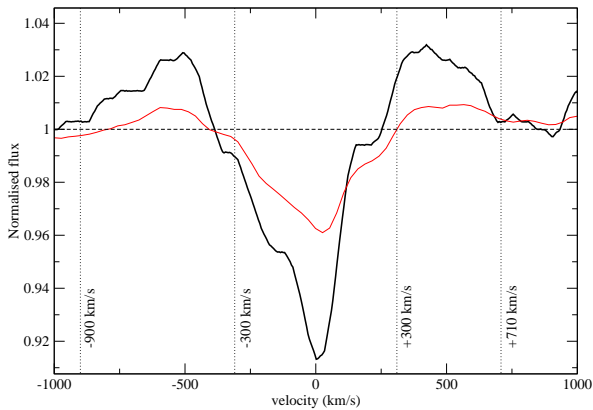}
\caption{\label{secLSD}He~{\sc i} $\lambda 4471$ line (thick black line) and LSD profile (thin red line) extracted from the disentangled spectrum of the secondary star. Vertical lines indicate $\pm v\sin i$ and the apparent extent of the wing emission.}
\end{figure}

Well-known examples of centrifugal magnetospheres exist amongst the magnetic B-type stars \citep[e.g. $\sigma$~Ori E;][]{2005ApJ...630L..81T,2012MNRAS.419..959O}. The prototypical optical spectral signature of these configurations is emission in Balmer lines at high velocities (i.e., several times $v\sin i$, as a consequence of magnetospheric plasma maintained in rigid rotation with the star, high above the stellar surface). In the case of $\sigma$~Ori E, for example, the emission extends to $\pm 600$~\kms, or about $\pm 3.75v\sin i$ \citep{2012MNRAS.419..959O}. (For reference, the Alfven and Kepler radii for $\sigma$~Ori E are estimated to be $1.9~R_*$ and $31.1~R_*$; Petit et al., in preparation.) In the case of HD 47129A2, analogous emission at high velocities is observed in many spectral lines \citep[principally lines of H and He~{\sc i}; see Figs. 1 and 8 of][]{2008A&A...489..713L}. Fig~\ref{secLSD} shows the LSD profile extracted from the disentangled spectrum of the secondary component, as well as the He~{\sc i} $\lambda 4471$ line from the disentangled spectrum. Vertical dotted lines indicate $\pm v\sin i$ ($=\pm 305$~\kms) and the estimated limits of the emission (at $-900$~\kms and $+710$~\kms). The emission therefore extends to approximately $2.3$ to $2.9$ times $v\sin i$. These values are certainly compatible with the Alfven and Kepler radii calculated above, and with the assumption (from Sect. 4.1) that the magnetic field in the emitting material contributes negligibly to the observed Stokes $V$ profile\footnote{Assuming a dipole field falling off as $1/r^3$ (as will be the case inside the Alfven radius), the field at these distances is 12-24 times weaker than at the stellar surface.}.

Attribution of this emission to a magnetosphere, rather than stellar wind collision, makes sense given that no similar emission is associated with the primary. In fact, \citet{2008A&A...489..713L} propose based on Doppler tomography that the secondary is surrounded by a rotationally-flattened wind, concentrated in the equatorial plane. They comment that this could explain the apparently stronger wind of the secondary (based on emission line strengths and X-ray absorption). It seems reasonable, given the detection of a magnetic field, to identify an equatorial enhancement of the secondary's wind with plasma confined by the star's magnetosphere. If the secondary's wind is structured into a magnetosphere, this may also help to explain the highly peculiar UV wind lines, in analogy with other magnetic early-type stars \citep[e.g.][]{2005ASPC..337..114H}.

The wind momentum ratio $\dot M_2 v_{\infty, {\rm 2}}/\dot M_1 v_{\infty, {\rm 1}}$ of the components is calculated to be 0.6 (using the measured wind properties of the primary) or 0.3 (using the computed wind properties of the primary). In both cases, we conclude that the primary's wind dominates. This implies that the primary's wind wraps around the secondary component, and places the wind collision at a distance of $3.1~R_*$ (or $1.6~R_*$) from the secondary. Thus the classical wind collision, in the absence of the dynamical influence of the magnetic field, is predicted to occur {\em inside the Alfven radius of the secondary component.} It could well be the primary's wind is actually deflected by the magnetic field of the secondary (just as the same magnetic field limits the wind of the secondary), in analogy to the deflection of the solar wind around planetary magnetospheres in the solar system. Or the primary's wind might be channelled along open field lines, effectively feeding the secondary's magnetosphere. In both scenarios the winds never reach a classical CWB collision, at least in a direction close to the line joining the two stars. Based on this reasoning, we should ask ourselves if the CWB scenario is in fact a good description of the HD 47129 system.

It is also interesting to consider that the magnetic field of the secondary is likely to be tilted relative to the orbit axis, and therefore that the magnetospheric structure will periodically rotate around or through the orbital plane, potentially generating a highly dynamic interaction with the wind of the primary. When the secondary's magnetic pole is directed at the primary, the relatively unimpeded wind of the secondary could produce a more typical colliding wind scenario.  But when the secondary presents its magnetic equatorial region to the primary, this should result in closed magnetic loops (and confined wind plasma of the secondary) being impacted by the primary wind. These differences could potentially influence the X-ray light curve.  Basically, one might expect modulation according to the rotation period of the secondary, with two peaks and troughs per rotation cycle, depending on geometry of the magnetic field (i.e. the angles $i$ and $\beta$).

\section{X-rays}

The properties of the X-ray emission of Plaskett's star reported
by \citet{2006MNRAS.370.1623L} based on their observations with \xmm have been interpreted as consistent with a CWB scenario, which
was reasonable given the knowledge of the system at that time. However,
the relatively high X-ray luminosity ($\log L_{\rm x} / L_{\rm bol} \sim -6.0$)
and temperature ($T_{max} \sim 2.5$~keV) are comparable to those of $\theta^1$~Ori C \citep{Gagne2005} and other known magnetic O stars, and could be
consistent with either MCWS or CWB scenarios.

We reanalyzed the archival \xmm RGS spectra of Plaskett's star
\citep{2006MNRAS.370.1623L} to obtain constraints on two key
parameters which could indicate the mechanism of its X-ray emission.
The ratio of the forbidden to intercombination lines ($f/i$) in
He-like triplets can constrain the proximity of the X-ray emitting
plasma to the stars \citep{Kahn2001}. Doppler broadening and shifts of
spectral lines can also provide contraints on the emitting geometry
\citep{OwockiCohen2001,Henley2003}.

We reprocessed the \xmm RGS data using SAS 12.0.0. We used {\tt
rgsproc} to produce spectra and response matrices, filtering intervals
of high background.

We fit the Si~{\sc xiii}, Mg~{\sc xi}, Ne~{\sc ix}, and O~{\sc vii}
He-like triplets with a three Gaussian model. We find that the
counting statistics of the data set are insufficient to provide
significant constraints on the $f/i$ ratio of any of these species.

We also fit several of the strongest lines with Gaussians to place
constraints on the Doppler broadening and shifts. The best constrained
line is O~{\sc viii} Ly$\alpha$, with best fit values given in Table~\ref{tab:xrayfit}.
To the extent that it can be constrained, the broadening is comparable to
that observed for O~{\sc viii} Ly$\alpha$ in the spectrum of $\theta^1$ Ori C.

Thus, we find that both CWB and MCWS models for X-ray emission are
consistent with the observed X-ray properties of Plaskett's
star. 

\begin{table}
\centering
 \caption{Fit results for O~{\sc viii} Ly$\alpha$. Uncertainties are reported for 95.3\% confidence interval.}
 \label{tab:xrayfit}
 \begin{tabular}{@{}lcc}
   \hline
   $\Phi$ & $\sigma^a$ & $\delta^b$ \\
             & (km s$^{-1})$ & (km s$^{-1}$) \\
   \hline
   0.1 & $<$ 1355 & $-355^{+700}_{-470}$ \\
   0.63 & $990^{+350}_{-330}$ & $205^{+400}_{-400}$ \\
   \hline\hline
 \end{tabular}
 \medskip
 \\
 $^a$Doppler broadening.\\
 $^b$Doppler shift.
\end{table}

\section{Discussion and conclusion}
\label{conclusion}

In this paper we have reported the firm detection of Zeeman signatures in high resolution Stokes $V$ spectra of the massive binary HD 47129 (Plaskett's star). We have demonstrated that the detected signatures are formed in the spectral lines of the rapidly rotating secondary component. No comparable field is observed in the sharper-lined primary. Based on the large inferred value of the longitudinal magnetic field, and its significant variation, we conclude that the field is highly organized over the star's surface. The surface polar strength of the dipole component of the field must be at least $2850\pm 500$~G. The field is observed to vary on timescales of days, compatible with rotational modulation of an oblique, static field in the rotating frame of the secondary star \citep[i.e. the oblique rotator model of][]{1950MNRAS.110..395S}. Although the data do not yet allow us to infer the period of variation of the longitudinal field, we find that the measurements do not phase coherently with the system's orbital period, nor with other periods derived from photometric or spectral variability. 

Computation of the magnetic wind confinement parameter $\eta_*$ using theoretical values of the secondary's mass loss rate and wind terminal velocity indicate that magnetic channelling should confine the wind of the secondary to a distance of $5.4~R_*$. Coupled with the observed rapid rotation, this confinement is expected to produce a long-lived centrifugal magnetosphere of stellar wind plasma in the region $2.2-5.4~R_*$. The high-velocity emission wings of the secondary's H and He~{\sc i} line profiles are quantitatively consistent with this scenario, while the highly peculiar UV wind profiles \citep{2008A&A...489..713L} of the secondary are qualitatively consistent. The X-ray spectrum observed with \xmm \citep{2006MNRAS.370.1623L} is also consistent with the expected equatorial shocks due to collision of the wind flows channelled by the magnetic field from opposite magnetic hemispheres. This suggests that previous interpretations of Plaskett's star as a classical CWB may be too simplistic. Indeed, the Doppler tomographic maps of He~{\sc ii} $\lambda 4686$ and H$\alpha$ computed by \citet{2008A&A...489..713L} are quite unusual for colliding wind systems in the sense that they reveal a ring of material in velocity space. Such ring-like structures are often seen in Algol-type binaries (and there they are attributed to the accretion disk). Since apparently neither of the stars in HD 47129 (currently) fills its Roche lobe, the most likely explanation (in view of the detection of the magnetic
field) is that we are seeing the emission from the magnetically confined wind. The line profile variations are also quite atypical for CWBs: most colliding wind systems where the bulk of the emission arises in the
wind-wind interaction zone display broad lines with changing profiles. In Plaskett's star, the broad component is relatively constant. What varies from one observation to the next is a narrow component that moves with some phase shift compared with the primary star. It is this narrow component that dominates the Doppler map; this is because Fourier filtering (as applied to the spectra in the Fourier filtered back-projection method used to build the Doppler map), actually removes the strictly constant part of the spectrum, i.e. in this case part of the confined wind emission). All of this suggests that what we are seeing in this
system is more complex than a classical wind-wind collision.



As mentioned in Sect. 1, Plaskett's star is proposed to be a post mass transfer (i.e. post RLOF) system \citep{1992ApJ...385..708B, 2008A&A...489..713L}. The RLOF scenario is invoked to explain, at least qualitatively, the mass-luminosity mismatch of the system, the peculiar surface chemistry of the components, and the secondary's rapid rotation. In this model, the primary was initially the more massive star. As it evolved, it filled its Roche lobe. The secondary gained rotational angular momentum during the mass transfer process, likely via an accretion disk located in the plane of the orbit. The secondary would only have needed to accrete a few percent of its own mass to reach critical rotation\footnote{Using the radius, mass and the maximum rotational period, we compute that the current rotation of the secondary corresponds to approximately 55\% of the critical rate (i.e. $\omega/\omega_{\rm crit} = 0.55$).} \citep{Packet1981}. During this event the angular momentum of the secondary would have been strongly modified, and likely forced into alignment with the axis of the orbit. { Rapid circularization probably also occurred during the RLOF event \citep[e.g.][]{2002MNRAS.329..897H}.}

The RLOF mass transfer process would certainly have represented a violent and transformative event in the life of the secondary component. If we imagine that the magnetic field we observe is a fossil field that existed prior to the RLOF event, we expect that mass transfer would have resulted in strong shear in the outer envelope, seriously distorting the structure of the field in the outer envelope. The shear would have acted on the poloidal component, generating toroidal field. If the toroidal field became sufficiently strong, and if the shear was not frozen by back-reaction of the field, a Tayler instability would likely have resulted. At the same time, Maxwell stresses and the Lorentz torque would have acted to damp differential rotation along field lines \citep[see][]{1988MNRAS.231..873M,2005A&A...440..653M} while the instability saturated.  Moreover, if the instability was was not able to regenerate both components of the initial field, the field would have decayed by ohmic diffusion. Such processes are observed by \citet{2007A&A...474..145Z}. Therefore, if a fossil field was present at the beginning \citep[e.g.][]{2006A&A...450.1077B,2010A&A...517A..58D}, its characteristics would have been strongly modified by the mass transfer implicit to the proposed RLOF. If this is indeed the case, it will be extremely interesting to investigate the detailed surface field properties, and to compare them to those of single O-type stars. 


A more speculative possibility is that the magnetic field was generated as a consequence of the RLOF event. While strong shear is expected to have been present during the mass transfer, it is not clear if the conditions required to drive a dynamo would have been satisfied. In this context, we note that \citet{2008MNRAS.387..897T} have proposed that highly magnetic white dwarfs have their origin in stellar mergers. The extension of this idea to explain the magnetic fields of non degenerate A, B and O type stars has been briefly described by \citet{2009MNRAS.400L..71F}. While the RLOF mass transfer and stellar mergers are very different events, the idea that highly dynamic and transformative events in the evolution of A, B and O stars might be the origin of their magnetic fields is provocative. Given that such events are now recognized to occur quite regularly in the Galaxy \citep[e.g.][]{2012Sci...337..444S}, these suggestions should certainly be investigated in more detail.

The uncertainties related to the age, distance, luminosities and chemical abundances of this system are a serious impediment to a more detailed understanding of its history and current evolutionary state. Resolution of some of these issues may need to await the Gaia mission. In the mean time, new observations in the optical and X-ray domains must be acquired. In particular, higher-SNR X-ray spectroscopy will constrain the X-ray emitting geometry and help to identify the emission mechanism. Phase-resolved spectroscopy over the entire optical domain will allow abundances of a large number of chemical elements - in particular tracers of mixing and mass transfer - to be determined for both components. Finally, additional polarimetric monitoring will allow the determination of the rotational period of the secondary star and the detailed geometry of its magnetic field.

Plaskett's star represents a heretofore unique stellar system - a close, massive binary with a rapidly rotating, magnetized component - that will be a rich system for further study. It will continue to be a priority target for MiMeS observations in the fall and winter of 2012, as well as a key focus of the new Binarity and Magnetic Interactions in various Classes of Stars (BinaMiCS) project and Large Program at the CFHT.

\section*{Acknowledgments}
GAW, JHG and AFJM acknowledge support from the Natural Sciences and Engineering Research Council of Canada (NSERC). SEdM acknowledges NASA Hubble Fellowship grant HST-HF-51270.01-A awarded by STScI, operated by AURA for NASA, contract NAS 5-26555. {\sc Clean-NG} was developed by B. Leroy at Observatoire Paris-Meudon, and kindly made available for this investigation. The authors acknowledge helpful input from J.Ma\'iz Apell\'aniz (Instituto de Astrofisica de Andalucia, Spain). CFHT and TBL observations were acquired thanks for generous allocations of observing time within the context of the MiMeS Large Programs.

\bibliography{Plaskett}

\end{document}